# Mechanisms of the magnetic incommensurability in *p*-type cuprate perovskites

A. Sherman

*Institute of Physics, University of Tartu, Riia 142, 51014 Tartu, Estonia*

phone: +372 737 4616

fax: +372 738 3033

e-mail: alexei@fi.tartu.ee

M. Schreiber

*Institut für Physik, Technische Universität Chemnitz, D-09107 Chemnitz, Germany*

**Abstract** We use the *t-J* model and Mori projection operator formalism for calculating the magnetic susceptibility of *p*-type cuprates in the superconducting and pseudogap phases. The lack of extended tails in the peaks of the hole spectral function was shown to provide an incommensurate low-frequency response with hole dispersions derived from photoemission. The theory reproduces the hourglass dispersion of the susceptibility maxima with the upper branch reflecting the dispersion of localized spin excitations and the lower branch being due to incommensurate maxima of their damping. The intensive resonance peak appears when the hourglass waist falls below the bottom of the electron-hole continuum. In the pseudogap phase, the Fermi arcs lead to a quasi-elastic incommensurate response for low temperatures. This result explains the lack of the superconducting gap in the susceptibility of phase-separated underdoped lanthanum cuprates. It may also explain the strengthening of the quasi-elastic response by magnetic fields and impurities. The theory accounts for the magnetic stripe reorientation from the axial to diagonal direction at low hole concentrations.

**Keywords** *p-type cuprates, magnetic properties, magnetic incommensurability, t-J model*

# 1. Introduction

Magnetic response of the $p$-type cuprate perovskites is incommensurate with a peculiar hourglass dispersion of the susceptibility maxima [1]. In this work we treat various aspects of the response in an approach, which takes into account strong electron correlations inherent in the crystals and does not presuppose charge stripes. The response differs essentially in the two low-temperature phases, the superconducting (SC) and the pseudogap (PG) states. This difference is one of the reasons of the variety of the observed magnetic properties.

We use the two-dimensional $t$-$J$ model, which is widely believed to describe holes and spins in Cu-O planes of cuprates. Applying the Mori projection operator technique the following general equation can be derived for the imaginary part of the magnetic susceptibility (an extensive consideration of this technique and the derivation of the below equations are given in [2]):

$$\chi''(\mathbf{k}\omega) = -\frac{\omega h_{\mathbf{k}} \operatorname{Im}\Pi(\mathbf{k}\omega)}{\left[\omega^2 - \omega \operatorname{Re}\Pi(\mathbf{k}\omega) - \omega_{\mathbf{k}}^2\right]^2 + \left[\omega \operatorname{Im}\Pi(\mathbf{k}\omega)\right]^2}, \tag{1}$$

where $h_{\mathbf{k}}$ is a slowly varying function near the antiferromagnetic momentum $\mathbf{Q} = (\pi, \pi)$, $\omega_{\mathbf{k}} \approx \sqrt{\omega_{\mathbf{Q}}^2 + c^2(\mathbf{k} - \mathbf{Q})^2}$ is the frequency of localized spin excitations with the gap $\omega_{\mathbf{Q}}$ defined by the correlation length of the short-range antiferromagnetic order. The polarization operator $\Pi(\mathbf{k}\omega)$ contains contributions of four processes, of which the decay of a spin excitation into an electron-hole pair dominates for small frequencies $\omega$. This contribution reads

$$\Pi(\mathbf{k}\omega) = -\sum_{\mathbf{k}'} f_{\mathbf{k}\mathbf{k}'}^2 \iint d\omega' d\omega'' \frac{N(\omega''\omega')\left[A(\mathbf{k}+\mathbf{k}',\omega'')A(\mathbf{k}'\omega') + A'(\mathbf{k}+\mathbf{k}',\omega'')A'(\mathbf{k}'\omega')\right]}{(\omega''-\omega')(\omega+\omega''-\omega'+i\eta)}, \tag{2}$$

where the interaction constant $f_{\mathbf{k}\mathbf{k}'}$ is a smooth function of its arguments, $\eta = +0$, $N(\omega''\omega') = n_F(\omega'') - n_F(\omega')$, $n_F(\omega) = 1/\left(e^{\omega/T} + 1\right)$ with the temperature $T$, $A(\mathbf{k}\omega)$ and $A'(\mathbf{k}\omega)$ are the hole normal and anomalous spectral functions, which have contributions from coherent and incoherent states. The function $A'(\mathbf{k}\omega)$ is proportional to the superconducting gap. Therefore, it is nonzero in the SC phase and vanishes in the PG phase.

## 2. Susceptibility in the superconducting phase

For the parameters of $p$-type cuprates above the frequency $\omega_r \approx \sqrt{\omega_\mathbf{Q}^2 + \omega_\mathbf{Q}\mathrm{Re}\Pi(\mathbf{Q}\omega_\mathbf{Q})}$ the momentum dependence of the susceptibility is defined by the denominator in (1). The respective susceptibility maxima form an upward-directed branch resembling the spin-wave dispersion, which has a gap $\omega_r$ at $\mathbf{k} = \mathbf{Q}$. In contrast to theories, in which the magnetic incommensurability is related to charge stripes (see, e.g., [3,4]), in the present approach the intensity of maxima is nearly isotropic for $\omega > \omega_r$, in agreement with experiment [1]. This branch of susceptibility maxima forms the upper part of the hourglass dispersion.

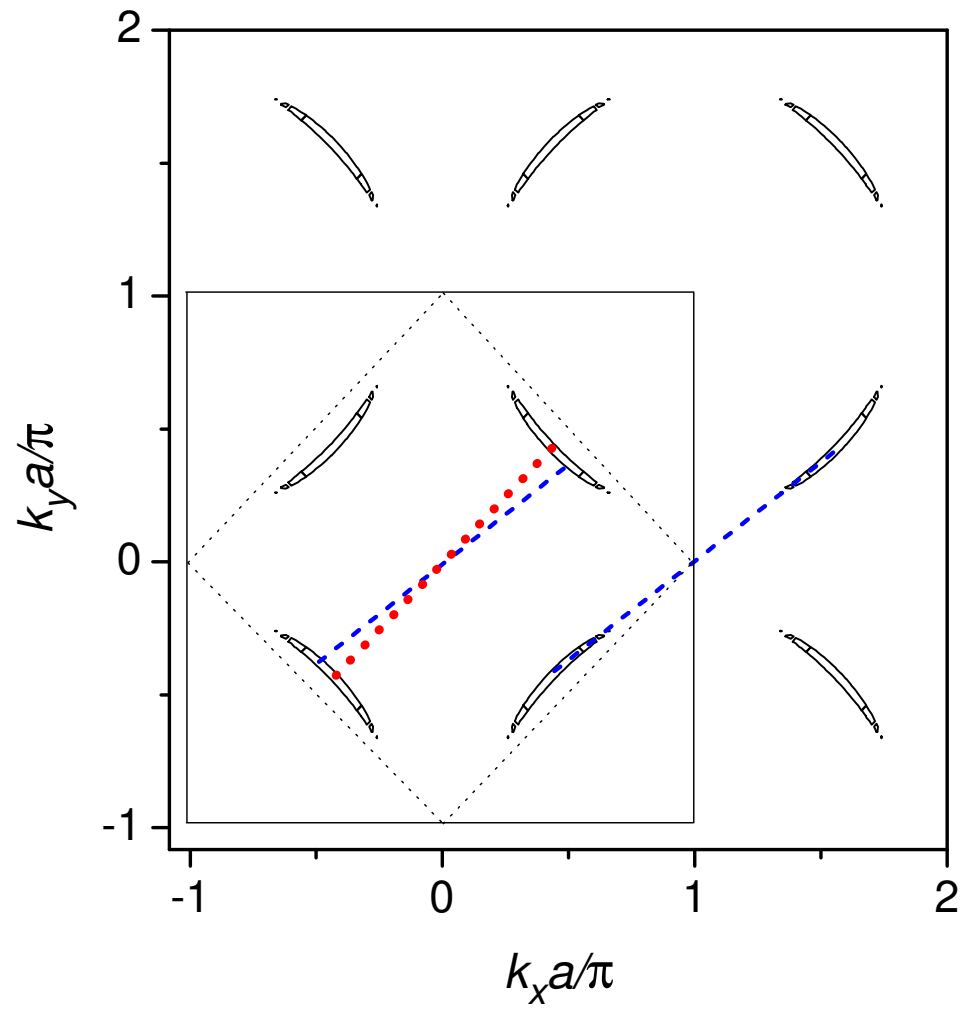

**Fig. 1**

As follows from (1), for small $\omega$ the behavior of $\chi''$ is determined by $\mathrm{Im}\Pi(\mathbf{k}\omega)$ in the numerator of the equation. Susceptibility maxima are connected with the coherent parts of the hole spectral functions in (2). As follows from this equation, in the SC phase fermions with energies $\xi_\mathbf{k} \approx \omega/2$ make the main contribution to $\mathrm{Im}\Pi(\mathbf{k}\omega)$. Here the SC dispersion $\xi_\mathbf{k} = \sqrt{\varepsilon_\mathbf{k}^2 + \Delta_\mathbf{k}^2}$ with the normal-state hole dispersion $\varepsilon_\mathbf{k}$ and the SC $d$-wave gap $\Delta_\mathbf{k}$. For small $\omega$ these states are located in crescent pockets near the boundary of the magnetic Brillouin zone (BZ, see Fig. 1). Due to peculiarities of $\varepsilon_\mathbf{k}$ in $p$-type cuprates the pockets settle inside this zone, in which opposite sides are displaced by the vector $\mathbf{Q}$ from each other.

Susceptibility maxima are related to the transitions between pockets (thick dashed and dotted lines in Fig. 1). Therefore, they appear at incommensurate positions. These maxima form the lower, descending branch of the hourglass dispersion.

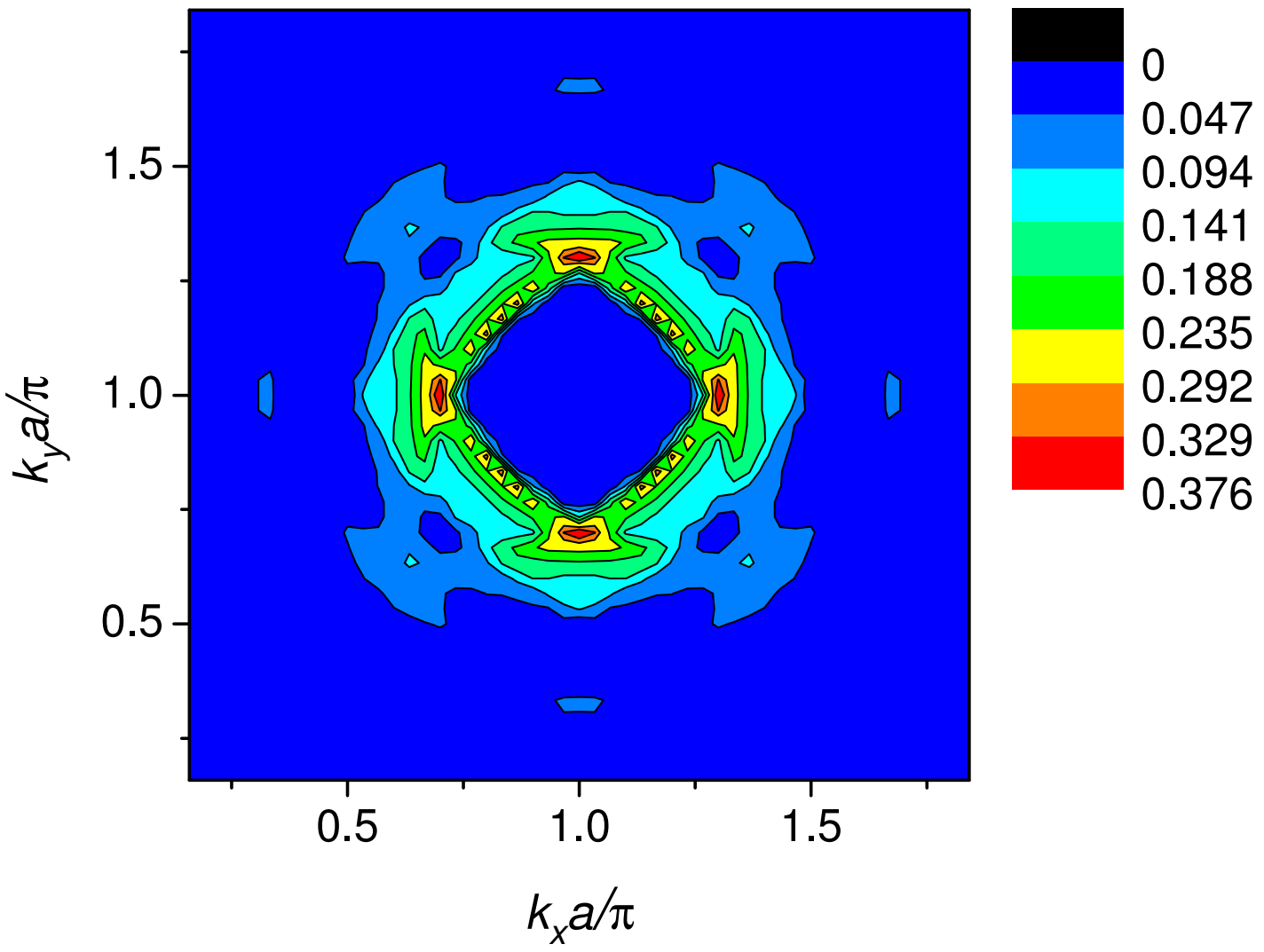

**Fig. 2**

The contour plot of $\chi''(\mathbf{k})$ calculated with the hole dispersion [5] and other parameters of $Bi_2Sr_2CaCu_2O_8$ for $\omega < \omega_r$ is shown in Fig. 2. The above-discussed mechanism of the formation of the maxima in $\chi''$ implies the fulfillment of the energy conservation $\omega = \xi_{\mathbf{k}'} + \xi_{\mathbf{k}'+\mathbf{k}}$. Due to long tails in Lorentzians usually used in the itinerant-electron theories for approximating coherent peaks in the hole spectral functions (see, e.g., [5,6]) this energy conservation can be violated, and the maxima are obtained only for $\varepsilon_{\mathbf{k}}$ with nested Fermi surfaces. However, the energy-independent damping in the Lorentzians contradicts the known properties of the hole self-energy in the *t-J* model. Therefore, the energy-conserving $\delta$-function in $\mathrm{Im}\Pi(\mathbf{k}\omega)$ was approximated with $[\theta(\omega+\Gamma)-\theta(\omega-\Gamma)]/2\Gamma$, $\Gamma$ being a broadening. In this case the hole dispersion derived [5] from photoemission was found to produce the low-frequency incommensurate response (Fig. 2) in good agreement with experiment [1]. Notice that in the itinerant-electron approach no incommensurability was found for small $\omega$ with this dispersion [5].

Due to the resonance denominator in (1) $\chi''(\mathbf{k}\omega)$ is peaked at $\omega \approx \omega_r$ and $\mathbf{k} = \mathbf{Q}$. This maximum is sharp and intensive, if $\omega_r$ falls into the region of the small damping $\omega_r \left|\mathrm{Im}\Pi(\mathbf{Q}\omega_r)\right|$ below the edge of the two-fermion continuum $\omega_e = \min(\xi_{\mathbf{q}} + \xi_{\mathbf{q}+\mathbf{Q}})$ at $\mathbf{Q}$ (see Fig. 3). Analogous maximum was observed in YBCO and some other cuprates and was called the resonance peak [7]. If $\omega_e < \omega_r$, the maximum falls into the region of a large damping [see Fig. 3(b)]. In this case the maximum becomes broader and less intensive, as observed in lanthanum cuprates [1,3].

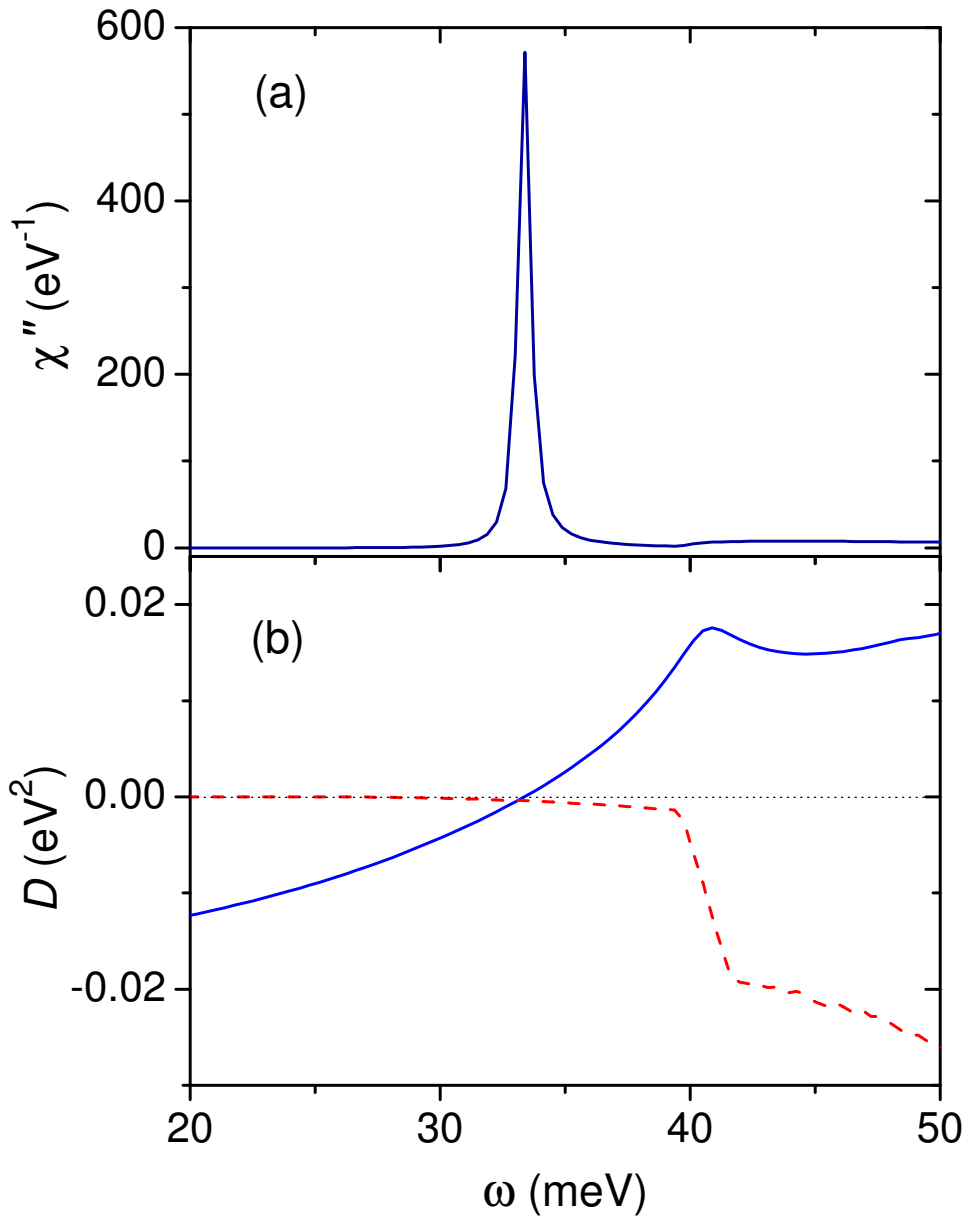

**Fig. 3**

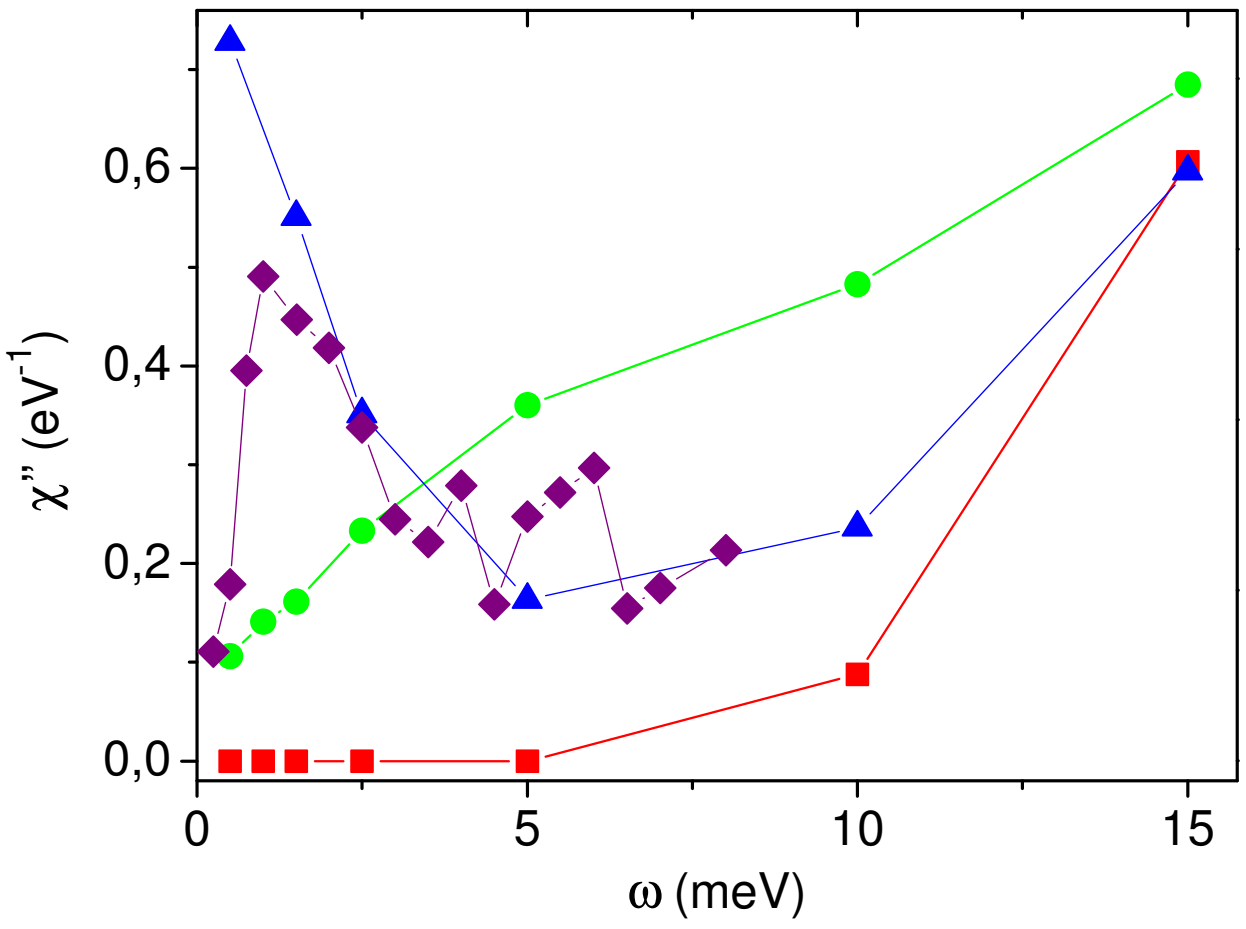

**Fig. 4**

For $\omega$ immediately below $\omega_r$ maxima of $\chi''$ are located on the BZ edge, since the number of fermion pairs contributing to $\mathrm{Im}\Pi(\mathbf{k}\omega)$ with such $\mathbf{k}$ is larger than for the diagonals (see Figs. 2 and 1, where there are two groups of transitions for a maximum on the edge and only one for a diagonal). With further decrease of $\omega$ in the SC phase the size of the crescent pockets diminishes and the intensity of the maxima drops. This leads to the SC gap in $\chi''$ (see Fig. 4, squares). The intensity on the BZ edge decreases faster with $\omega$ than on the diagonals, since the fermion states contributing to $\mathrm{Im}\Pi(\mathbf{k}\omega)$ are located on the periphery of the pockets in the former case and near their centers in the latter (see Fig. 1). Therefore, with decreasing $\omega$ the maxima of $\chi''$ transfer from the BZ edge to the diagonals.

## 3. Susceptibility in the pseudogap phase

This behavior of $\chi''$ is changed in the PG phase, which is characterized by Fermi arcs – fragments of the Fermi surface near the node points [8]. They resemble crescent pockets in Fig. 1; however, in contrast to them in the PG phase the length of the equi-energy contours remains finite with $\omega \to 0$. As a consequence $\chi''$ does not vanish down to $\omega \sim T$ (see Fig. 4, circles). Moreover, in the PG phase $\chi''$ can grow with decreasing $\omega$ (triangles in Fig. 4). The reason for this growth is a small cusp in the hole dispersion at the Fermi level, which was obtained in the self-consistent calculations in the *t-J* model [9], and which was presumably observed in photoemission [8]. The cusp increases the combined density of states at small $\omega$, which leads to the growth of $\chi''$. This behavior resembles that observed [10] in underdoped lanthanum cuprates (Fig. 4, diamonds). Thus, in contrast to the SC phase with the SC gap in the susceptibility, the magnetic response of the PG phase is quasi-elastic [11].

In the SC phase, the underdoped cuprates differ from more heavily doped crystals by the lack of the SC gap in the susceptibility, which remains finite down to low frequencies [1]. The results shown in Fig. 4 can account for this distinction, if additionally a phase separation into SC and PG regions is supposed [9] in the former compounds. In these crystals, the SC gap is not observed because the low-frequency response is determined by the PG regions, while both types of regions contribute to $\chi''$ for larger $\omega$. Thereby the incommensurability parameter varies continuously from small to large $\omega$ and the dispersion of maxima has the hourglass shape, as in more heavily doped cuprates. The same mechanism may be responsible for the enhancement of incommensurate quasi-elastic peaks of $\chi''$ by modest magnetic fields [12] and small impurity concentrations (Fujita, M., Enoki, M., Iikubo, S, Kudo, K, Kobayashi, N., Yamada, K.: arXiv:0903.5391). An external magnetic field, which is smaller than the upper critical field, produces vertices in the crystal. In accord with some theories and experimental data [13,14] vertex cores and their neighborhood are in the PG phase. Thus, in the underdoped case the field increases the fraction of the crystal occupied by this phase, thereby strengthening the quasi-elastic signal, as observed in [12]. In optimally doped LSCO vertices induced by the field lead to the partial filling of the SC gap [15]. Regions of the PG phase around impurities furnish the same result.

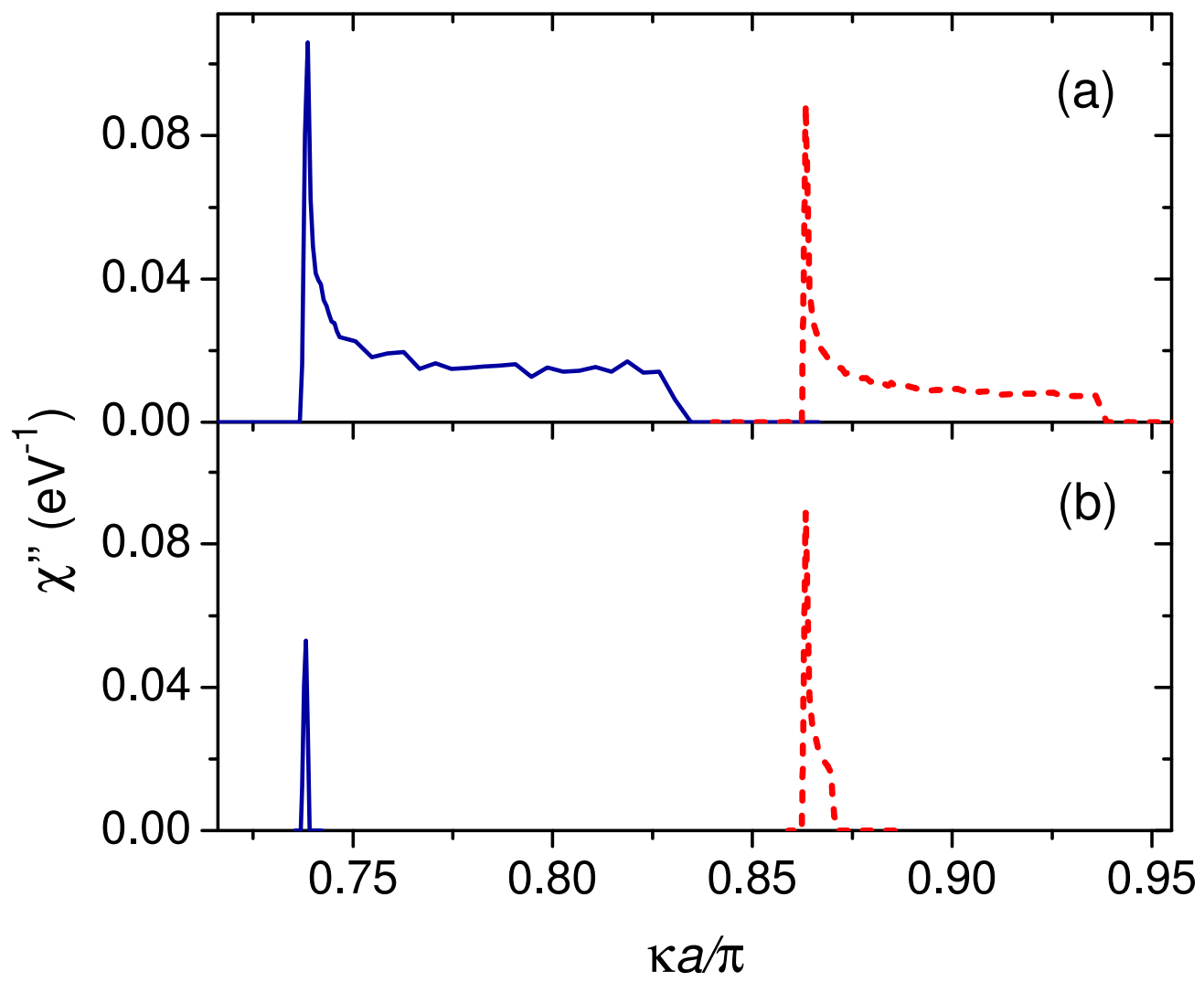

**Fig. 5**

The above results allow us to propose the mechanism of the reorientation of the low-frequency susceptibility maxima from the axial to the diagonal direction, which is observed in the PG phase of LSCO for $x \leq 0.05$ [1]. For large arc lengths $\chi''$ is peaked at the BZ edge down to $\omega \sim T$ [see Fig. 5(a)], since in this case the number of fermion pairs contributing to the spin-excitation damping is larger for such momenta than for diagonals (the analogous situation for the SC phase is shown in Fig. 1). However, the arc length decreases nearly linearly with $x$ [8]. As in the SC phase, the reduction of the length of the equi-energy contours suppresses $\chi''$ on the BZ edge more strongly than on the diagonal [see Fig. 5(b)]. For the parameters of Fig. 5 $\chi''$ becomes peaked on the BZ diagonal at $l \approx 0.6a^{-1}$, which corresponds to $x \approx 0.04$ [8]. This value is close to the mentioned concentration of the maxima reorientation in LSCO. Notice that also for such small $x$ the dispersion of the susceptibility maxima has the hourglass shape. However, the values of $\omega_r$ and the incommensurability parameter are smaller than for larger $x$. These results agree with experimental observations [1].

The theory based on Mori's formalism [2,11] takes proper account of strong electron correlations, which are inherent in cuprates. In this respect, this theory differs from weak-interaction itinerant-electron approaches [5,6]. In contrast to them, for the description of the incommensurate response the strong-correlation theory does not need in the Fermi surface nesting or in vanishingly small damping of hole states. Besides, this theory describes the magnetic response of $n$-type cuprates (Sherman, A.: arXiv:1207.3405), while the itinerant-electron approach has difficulties in its interpretation [16]. In contrast to works [3,4] the present theory does not need in the supposition of static or quasi-static charge stripes, which existence was safely detected only in some of the *p*-type cuprates [17]. The stripe theory does not take into account the difference in the hole dispersion in SC and PG phases.

**Acknowledgements** This work was supported by the European Union through the European Regional Development Fund (project TK114) and by the Estonian Science Foundation (grant ETF9371).

Fig. 1 The contour plot of the SC dispersion $\xi_{\mathbf{k}} = \sqrt{\varepsilon_{\mathbf{k}}^2 + \Delta_{\mathbf{k}}^2}$. The contours correspond to the energy 15 meV. The squares shown by thin solid and dotted lines are the first and magnetic BZ. The thick dashed and dotted lines connect fermion pairs contributing to $\chi''$ on the BZ edge and diagonal, respectively. The lattice spacing is denoted by $a$.

Fig. 2 The contour plot of $\chi''(\mathbf{k})$ calculated [2] with the hole dispersion derived [5] from the photoemission data in $Bi_2Sr_2CaCu_2O_8$. The frequency $\omega = 30$ meV, the hole concentration $x \approx 0.17$.

Fig. 3 The resonance peak (a), and the real (solid line) and imaginary (dashed line) parts of the denominator in $\chi$ (b). The edge of the two-fermion continuum is seen as a step at $\omega \approx 40$ meV. Here the exchange constant $J = 0.1$ eV, the hole nearest and next-nearest neighbor hopping constants $t = 0.5$ eV and $t' = -0.15$ eV, respectively, the hole concentration $x = 0.12$, and the SC gap $\Delta = 25$ meV.

Fig. 4 The maximum susceptibility on the BZ edge in the SC (squares) and PG (circles) phases at $T = 0$. Triangles show $\chi''$ in the PG phase in the presence of a cusp in the hole dispersion at the Fermi level. For the SC phase parameters are the same as in Fig. 3. In the PG phase the arc length $l = 2a^{-1}$, which corresponds to $x \approx 0.12$ [8], and the PG magnitude $\Delta_{PG} = 25$ meV. The susceptibility measured [10] in $La_{1.875}Sr_{0.125}CuO_4$ at $T = 4$ K is shown by diamonds (in arbitrary units).

Fig.5 The momentum dependence of $\chi''$ for $\omega = 0.5$ meV along the BZ edge [solid lines, $\mathbf{k} = (\pi/a, \kappa)$] and along the diagonal [dashed lines, $\mathbf{k} = (\kappa, \kappa)$]. The length of the Fermi arcs $l = 2a^{-1}$ in panel (a) and $l = 0.6a^{-1}$ in panel (b).